\documentclass[preprint,11pt]{revtex4}
\usepackage{latexsym}

\usepackage{graphicx}

\begin{document}

\title{Stretching of a chain polymer adsorbed at a surface.}

\author{J.\ Krawczyk}
\email{krawczyk.jaroslaw@tu-clausthal.de}
\author{T.\ Prellberg}
\email{thomas.prellberg@tu-clausthal.de}
\affiliation{Institut f\"ur Theoretische Physik,
Technische Universit\"at Clausthal,
Arnold Sommerfeld Stra\ss e 6,
D-36578 Clausthal-Zellerfeld,
Germany}
\author{A.\ L.\ Owczarek}
\email{aleks@ms.unimelb.edu.au}
\author{A.\ Rechnitzer}
\email{andrewr@ms.unimelb.edu.au}
\affiliation{Department of Mathematics and Statistics, The University of Melbourne, 3010, Australia}

\begin{abstract}
In this paper we present simulations of a surface-adsorbed polymer
subject to an elongation force. The polymer is modelled by a
self-avoiding walk on a regular lattice.  It is confined to a
half-space by an adsorbing surface with attractions for every vertex
of the walk visiting the surface, and the last vertex is pulled
perpendicular to the surface by a force.  Using the recently proposed
flatPERM algorithm, we calculate the phase diagram for a vast range of
temperatures and forces.  The strength of this algorithm is that it
computes the complete density of states from one single simulation.
We simulate systems of sizes up to $256$ steps.
\end{abstract}


\maketitle
\section{Introduction}

New experimental methods in the physics of macromolecules
\cite{strick2001} have been used to study and manipulate single
molecules and their interactions.  These methods make a contribution
to our understanding of such phenomena as protein folding or unzipping
DNA; one can push or pull a single molecule and watch how it responds.
It is possible to apply (and measure) forces large enough to induce
structural deformation of single molecules.  One can monitor the
mechanism of some force-driven phase transition occurring at the level
of a single molecule. Theoretical understanding of this behaviour has
attracted much attention
\cite{marenduzzo2003,rosa2003,orlandini2004}.  New features are
observed if one pulls a macromolecule localised near an adsorbing
surface \cite{vrbova1996}.  One observes two phases: an adsorbed phase
and a desorbed phase. The desorbed phase is characterised by the mean
fraction of molecules in the adsorbing plane going to zero as the
number of molecules in the chain goes to infinity.  For a given
temperature one can find the critical force at which the
macromolecules are desorbed.  The phase-diagram in the
force-temperature plane can show re-entrant behaviour similar to that
found in DNA unzipping models \cite{orlandini2001} and directed walk
models \cite{orlandini2004}. 
  
Lattice models play an important role in the study of equilibrium
properties of linear polymer molecules.  Including interactions
between monomers and a surface confining the polymer, it is possible
to investigate phenomena such as the adsorption-desorption transition.
The pulling of directed polymers is already well investigated and
understood \cite{orlandini2004,rosa2003}.  We use self-avoiding walks
(SAW) on a regular lattice to study the adsorption of a polymer at a
surface subject to an elongation force. Vrbova and Whittington studied
the phase diagram for adsorbing interacting self-avoiding walks using
rigorous arguments \cite{vrbova1996} and simulations with the Markov
chain method employing pivot steps \cite{vrbova1998}.  The transition
studied by them for polymers in a good solvent (without interaction
between monomers) is equivalent to temperature-driven adsorption
(without force) in our SAW-model. The model of SAW for force-induced
desorption was already investigated by Mishra {\it et.al.}
\cite{mishra2004} using exact enumeration, which gave the correct
phase diagram but due the rather small system sizes studied, the
location of the phase boundary for infinite systems was not very
precise.  In this paper we present an investigation of this problem
using a new flatPERM algorithm \cite{prellberg2004}.  This is a very
good tool to easily get information of the whole phase diagram by
calculating the complete density of states.  While the exact
enumeration study \cite{mishra2004} was restricted to $n\le19$ steps,
flatPERM allowed us to perform ``approximate'' enumeration up to
$n=256$ steps.

\begin{figure}[t]
\includegraphics[scale=0.7,angle=0]{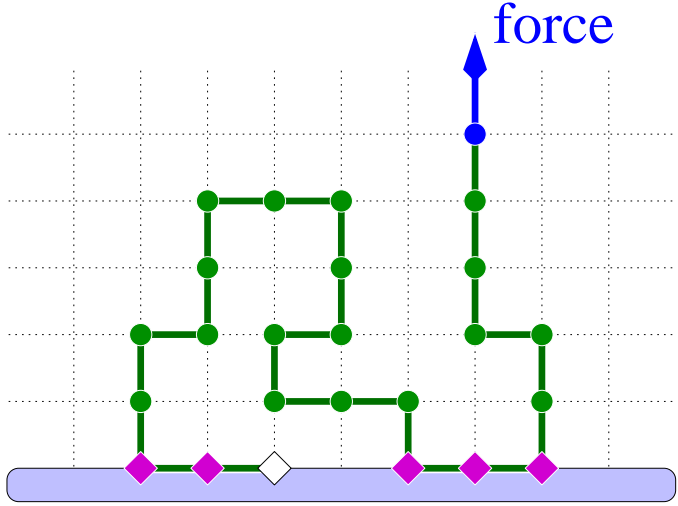}
\caption{\label{conf_ex} Example of  a configuration near a
surface. Monomers interacting with the surface (in two dimensions this
is a line) are denoted by diamonds. The first vertex, which is fixed
to the surface, is denoted by a white-filled diamond. The elongation
force acts only at the last monomer pulling it in the perpendicular
direction to the surface.}
\end{figure}

\section{Definition of the model}
\label{model}

We consider a self-avoiding walk on the simple cubic and square
lattices confined to the half-space or half-plane, $z\ge0$ or $y\ge0$,
respectively.  We define a {\it visit} as a vertex, representing a
monomer, lying in the surface $z=0$ in three dimensions or on the line
$y=0$ in two dimensions. The monomers interact with the surface via an
interaction strength, $\epsilon=-1$. In addition we have an elongation
force $f$ acting on the last monomer, pulling it away from the
surface, i.e.\ in the positive $z$-direction or positive
$y$-direction, in three and two dimensions respectively; see Fig.\
\ref{conf_ex}.  Therefore we have two competing effects: attraction to
the surface leading to adsorption and the elongation force leading to
desorption.  The partition function of the model is given by
\begin{equation}
\label{eq_part_fn}
Z_n(\omega_s,\omega_f)=\sum_{m_s,h}C_{n,m_s,h}\; \omega_s^{m_s}\omega_h^h,
\end{equation}
where $C_{n,m_s,h}$ is the number of all configurations with $n+1$
vertices (monomers) with one end at some fixed origin at the surface
$z=0$ or line $y=0$, respectively.  The number of visits (including
the fixed site) is denoted by $m_s$, and $h$ is the distance of the
($n+1$)st vertex from the surface.  The Boltzmann weight
$\omega_s=e^{-\beta\epsilon}=e^{\beta}$ ($\epsilon=-1$) is associated
with the interaction with the surface and $\omega_h=e^{\beta f}$ with
the elongation force $f$, where $\beta=1/k_BT$. We define a
finite-size free energy $\kappa_n(\omega_s,\omega_h)$ per step as
\begin{equation}
\kappa_n(\omega_s,\omega_h)=\frac1n\log Z_n(\omega_s,\omega_h).
\end{equation}
The usual free energy is related to this by $-\beta F_n\equiv
n\kappa_n(\omega_s,\omega_h)$.  In our simulation we obtain estimates
of $C_{n,m_s,h}$, so that a quantity $Q_{n,m_s,h}$ averaged over the set
of parameters $(m_s,h)$ for a given length $n$ is calculated by
\begin{equation}
\label{eq_quant}
\left<Q\right>_n(\omega_s,\omega_f)=\frac{\sum\limits_{m_s,h}Q_{n,m_s,h}C_{n,m_s,h}
\omega_s^{m_s}\omega_h^h}{\sum\limits_{m_s,h}C_{n,m_s,h}\omega_s^{m_s}\omega_h^h}.
\end{equation}
In this paper we concentrate on the adsorption and elongation of the
self-avoiding walk and the corresponding phase diagram.  We
investigate the behaviour of the average distance of the last monomer
from the adsorbing surface
\begin{equation}
\label{eq_h}
\left<h\right>=n\frac{\partial\kappa_n}{\partial\log\omega_h},
\end{equation}
the average numbers of monomers interacting with the surface 
\begin{equation}
\label{eq_ms}
\left<m_s\right>=n\frac{\partial\kappa_n}{\partial\log\omega_s},
\end{equation}
and the fluctuations of $m_s$ 
\begin{equation}
\label{eq_flkt}
\sigma^2(m_s)=\langle m_s^2\rangle-\langle m_s\rangle^2=n
\frac{\partial^2\kappa_n}{\partial^2\log\omega_s}.
\end{equation}

Since the desorption transition is characterised by a significant
change in the number of surface adsorbed sites, we  also investigate
the fluctuations of $m_s$ as a signature of the transition.  We are
interested in determining the location of the adsorption transition
for the whole range of forces and temperatures.

\section{Algorithm}
\label{algorithm}
Since we are interested in investigating the complete phase diagram,
one needs to perform simulations for the whole range of temperatures
and forces.  Conventionally, one would carry out different simulations
for numerous values of temperature and force to investigate the region
of interest.  With the flatPERM algorithm it is possible to cover the
whole range (given sufficient time for the simulation to converge)
with one single simulation. The flatPERM algorithm is a recently
proposed stochastic growth algorithm
\cite{prellberg2004}, which performs an estimation of the whole
density of states and can be interpreted as an approximate counting
algorithm.  The algorithm combines the pruned-enriched Rosenbluth
method (PERM) \cite{grassberger1997} with umbrella sampling techniques
\cite{torrie1977}. The configurations of interest are grown from
scratch adding a new monomer at each step.  We parameterise the
configuration space in such a manner that the algorithm explores it
evenly; i.e.\ for every set of parameters $(n,m_s,h)$ it aims to
generate the same number of samples. This requirement leads to a flat
histogram in the parameterisation.  Here we choose as parameters the
surface energy (number of contacts, $m_s$) and the distance of the
last monomer from the adsorbing surface, $h$.  During one simulation
we are able to explore all possible sets of parameters (all vectors
($n,m_s,h$) for all $n\leq 256$) and estimate the associated density
of states $C_{n,m_s,h}$.  As an example, for $n=256$ we have
calculated a histogram over $33151$ vectors for both dimensions.

\begin{figure}[ht!]
\includegraphics[scale=0.9,angle=0]{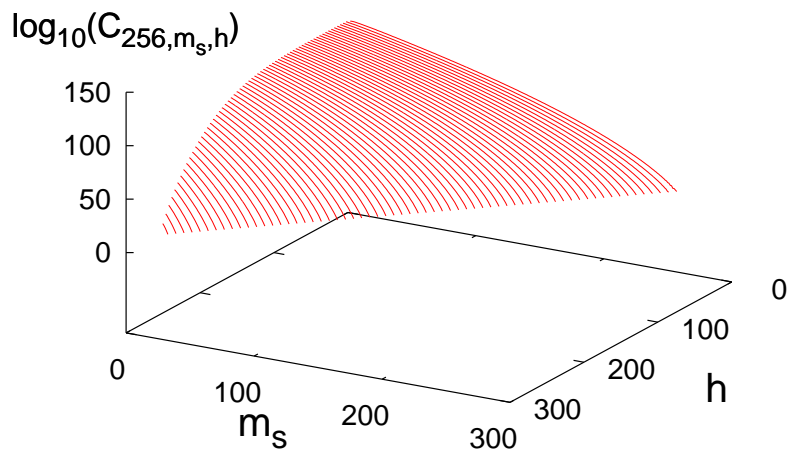}
\includegraphics[scale=0.9,angle=0]{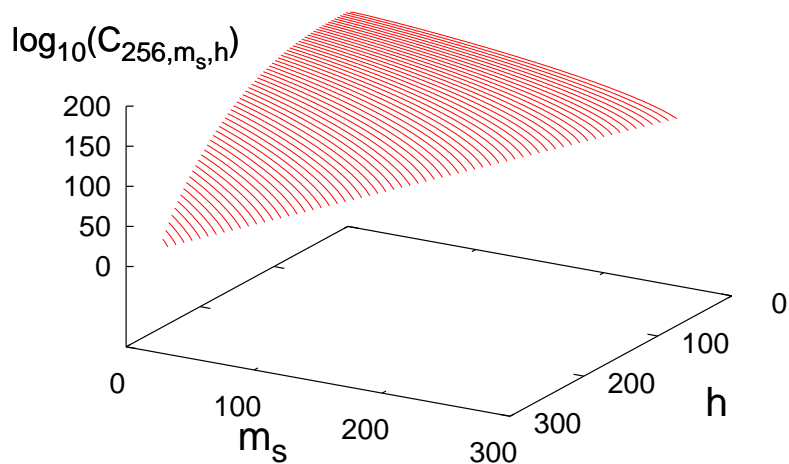}
\caption{\label{density}  Density of states for $n=256$ in two (top)
and three (bottom) dimensions. The weights of all vectors
($256,m_s,h$) were found during one simulation with flatPERM.}
\end{figure}

\section{Results}
\label{results}
In this section we present the results for both two and three
dimensions.  Fig.\ \ref{density} shows the density of states for two
and three dimensions for polymers of length $n=256$ steps.  From this
we can calculate all quantities of interest using eqn.\
(\ref{eq_quant}). Because we are focusing on the transition between
desorbed and adsorbed phases we consider the changes in both $m_s$ and
$h$. 
\begin{figure}[hb!]
\includegraphics[scale=0.9,angle=0]{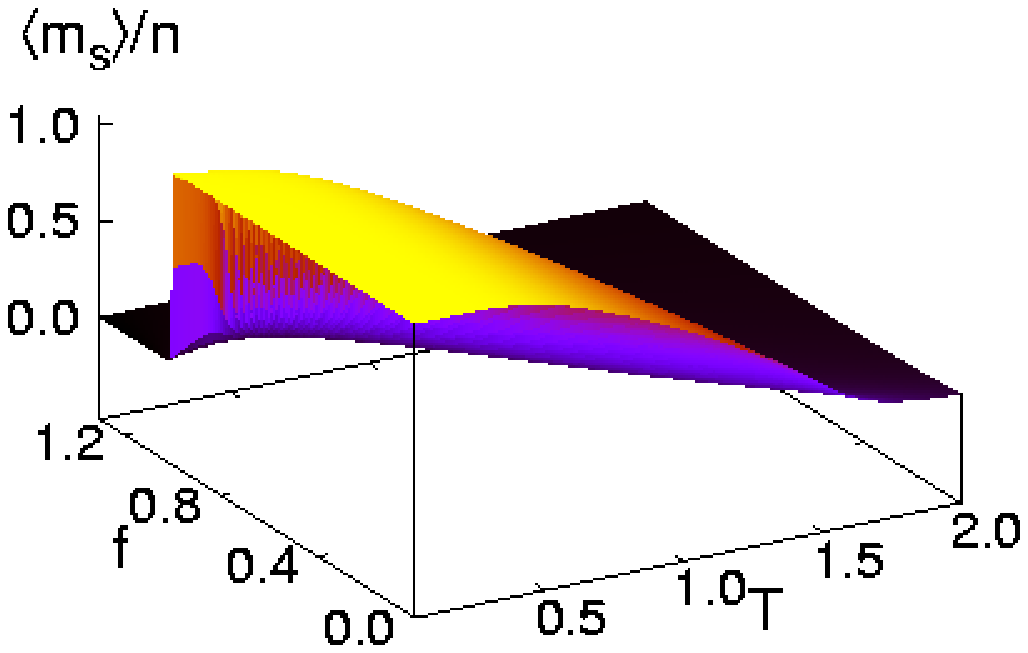}
\includegraphics[scale=0.9,angle=0]{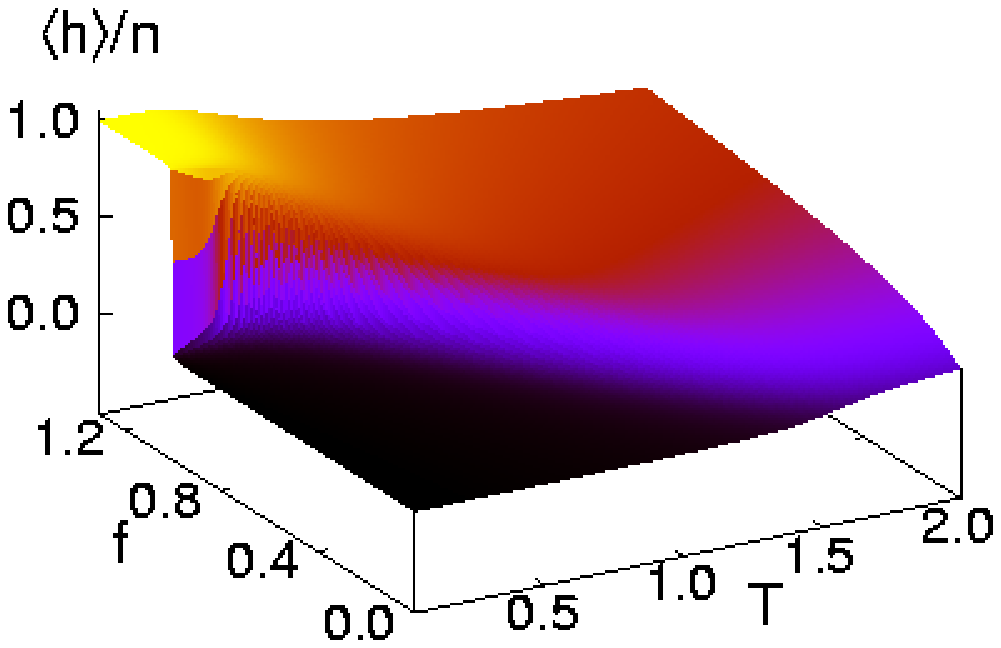}
\caption{\label{h_ms_2D} The average number of surface contacts $m_s$
(top) and the average height per monomer $h/n$ (bottom) in two
dimensions at length $n=256$. One can see two well-distinct phases.
The desorbed phase is characterised by $h>0$ and $m_s\approx 1$. For
the adsorbed phase $m_s$ reaches its maximal value while $h\approx
0$. If the force is bigger than one the system is desorbed for all
temperatures.}
\end{figure}
\begin{figure}[ht]
\includegraphics[scale=0.9,angle=0]{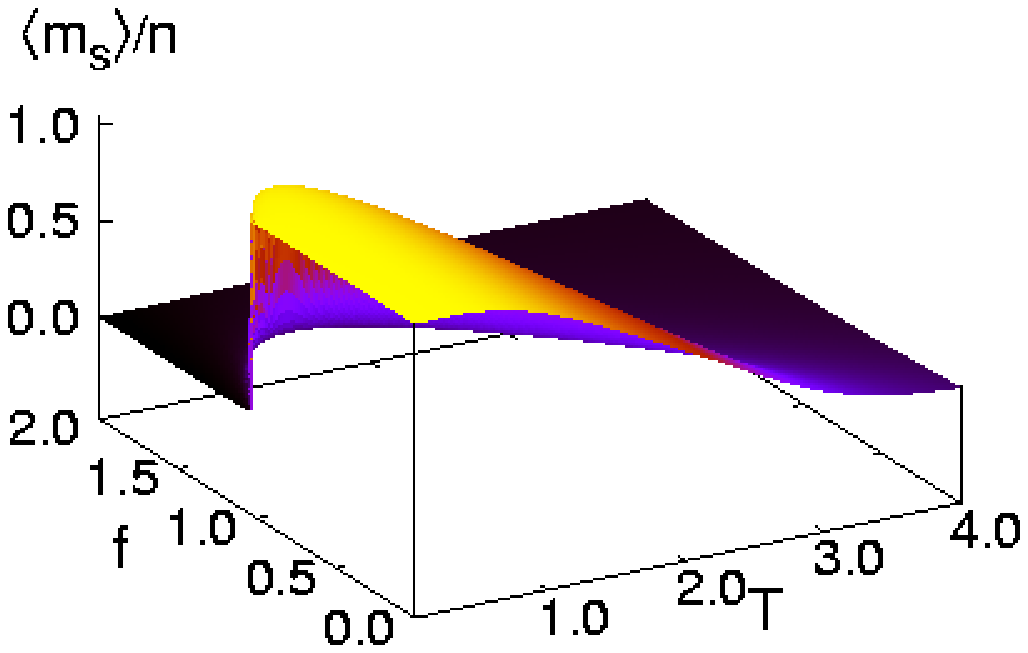}
\includegraphics[scale=0.9,angle=0]{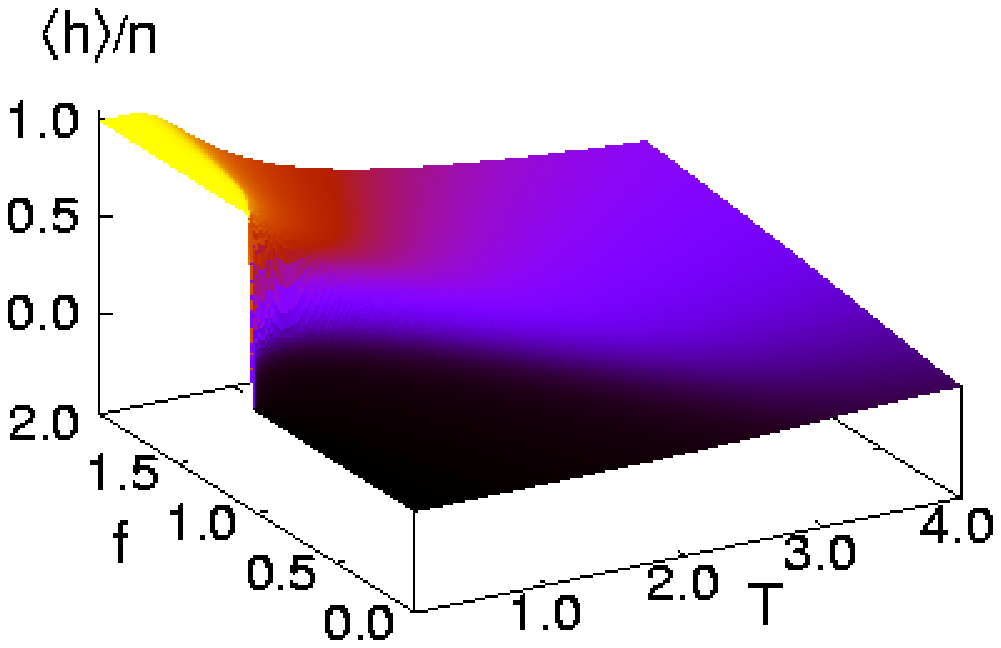}
\caption{\label{h_ms_3D}  The average number of surface contacts $m_s$
(top) and the average height per monomer $h/n$ (bottom) in three
dimensions at length $n=256$. For values of force $f < 1$ the
behaviour of the system is similar to the two-dimensional case moving
from a desorbed phase at low temperatures to a desorbed phase for high
temperatures. Above the critical force $f=1$ re-entrant behaviour
occurs up to some maximum force $f_{max}$. For some force $f$, with
$f_{max}\geq f >1$, the system moves from the desorbed to adsorbed and
back to the desorbed phase as the temperature is increased from near
zero.  }
\end{figure}

At low temperatures we find, for both dimensions, a clear
indication of a desorption transition between an adsorbed state in
which the average number of surface contact is maximal ($m_s\approx
n$) to an elongated desorbed state in which the polymer is completely
stretched ($h\approx n$) and pulled away from the surface. For higher
temperatures this transition persists up to a critical temperature, at
which thermal fluctuations alone lead to desorption.  In Fig.\
\ref{h_ms_2D} one can see the behaviour of the average number of
visits $m_s$ and the average distance $h$ of the last vertex from the
adsorbing surface for two dimensions, $n=256$.  For forces $f>1$ in
two dimensions the system is desorbed for all temperatures.  In three
dimensions, Fig.\
\ref{h_ms_3D} shows that the behaviour is similar but not identical.
Here we see the re-entrant behaviour previously observed in directed
models \cite{orlandini2004}. For a range of forces $f>1$, though not
too large, if one fixes the force and considers going from small to
high temperatures the system is first desorbed then at some
temperature depending on the force becomes adsorbed, and after further
increasing the temperature further the system becomes once again
desorbed. Such behaviour does not appear in two dimensions.

\begin{figure}[ht!]
\includegraphics[scale=0.9,angle=0]{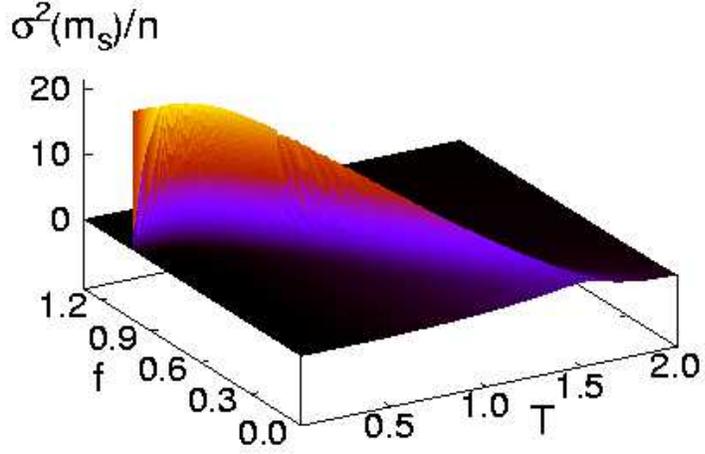}
\includegraphics[scale=0.9,angle=0]{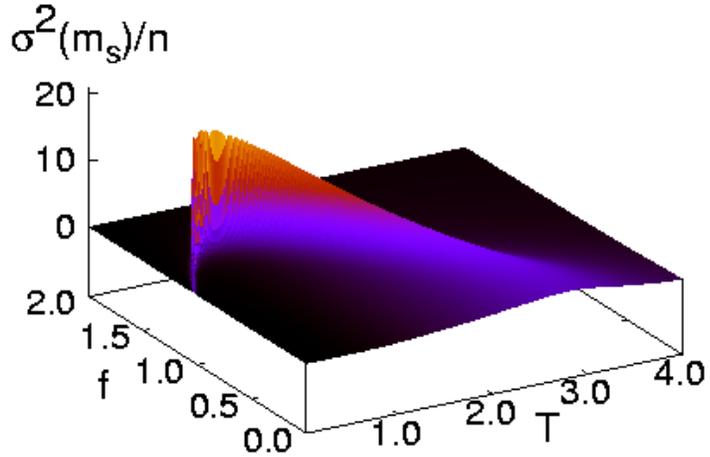}
\caption{\label{flukt_ms} Fluctuations of $m_s$ in two (left) and
three (right) dimensions for $n=256$. One can distinguish two
different phases (an adsorbed phase and a desorbed phase), which are
separated by a peak in the fluctuations.}
\end{figure}
\begin{figure}[ht!]
\includegraphics[scale=0.9,angle=0]{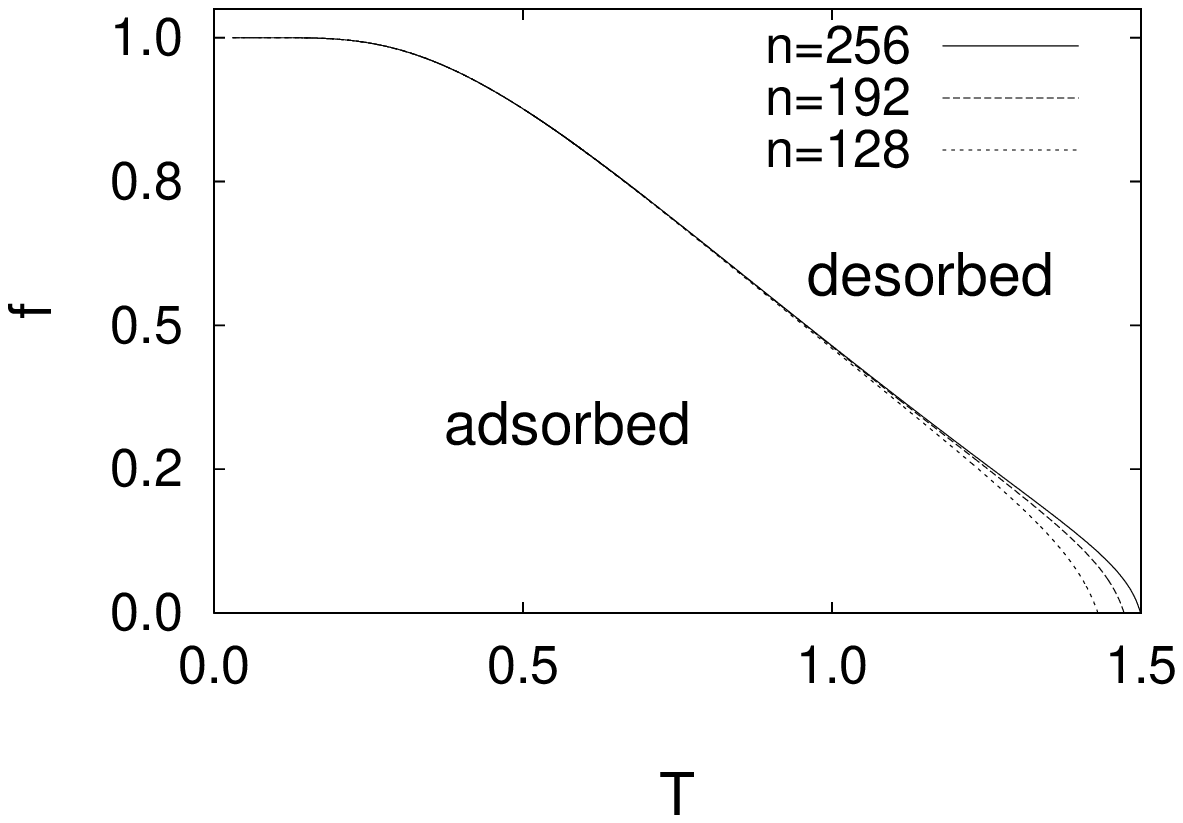}
\includegraphics[scale=0.9,angle=0]{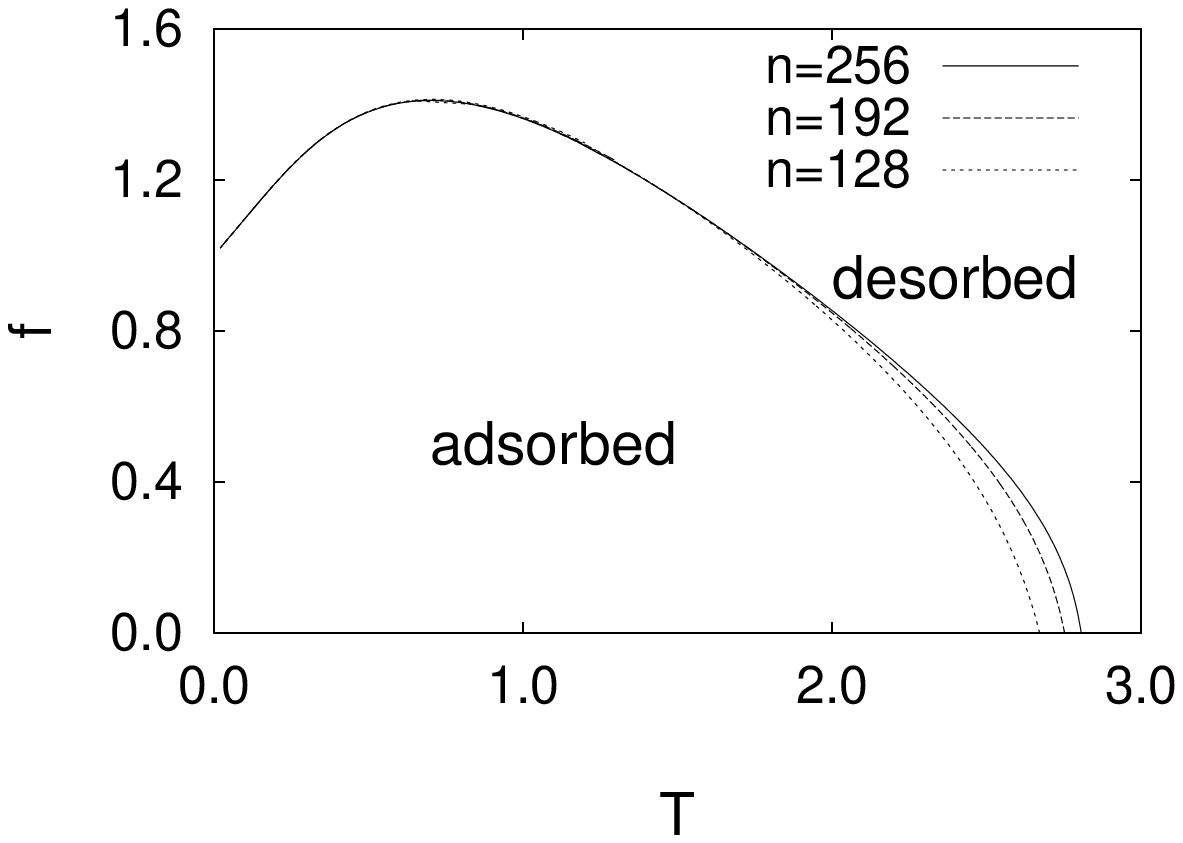}
\caption{\label{phase_diagram} Phase diagram for three different
lengths for two (top) and three (bottom) dimensions. One can clearly
see the difference between the systems. In three dimensions we have
clear re-entrant behaviour. There is no difference in the transition
position for small temperature and high force, while for higher
temperature and small force the position of the transition depends on
the system size $n$.}
\end{figure}

We estimate the phase boundary looking for positions of maximal
fluctuations in $m_s$.  The fluctuations in $m_s$ for $n=256$ are
shown for both dimensions in Fig.\ \ref{flukt_ms}. The fluctuations
separate two distinct phases (the adsorbed phase and the desorbed
phase).  The whole phase diagram for three different sizes of the
system is shown in Fig.\ \ref{phase_diagram}.

\begin{figure}[ht]
\includegraphics[scale=0.9,angle=0]{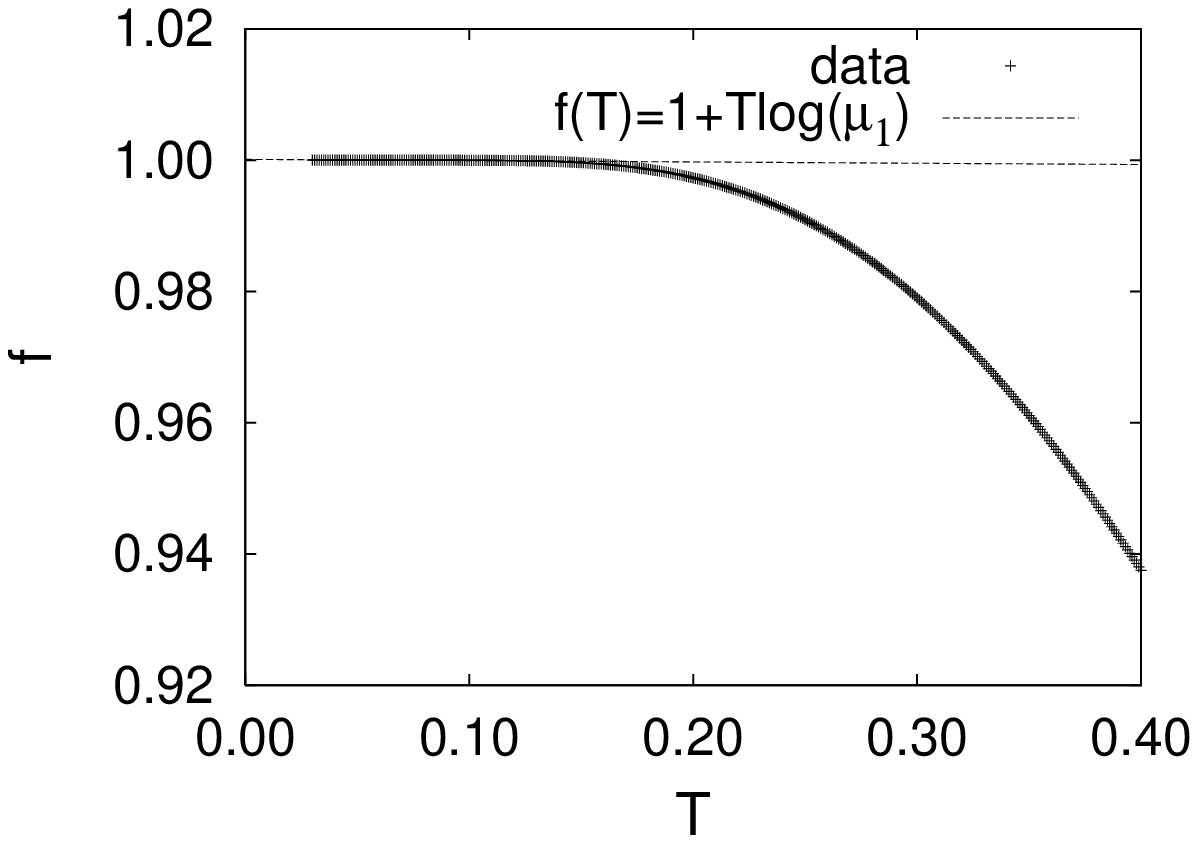}\hfil
\includegraphics[scale=0.9,angle=0]{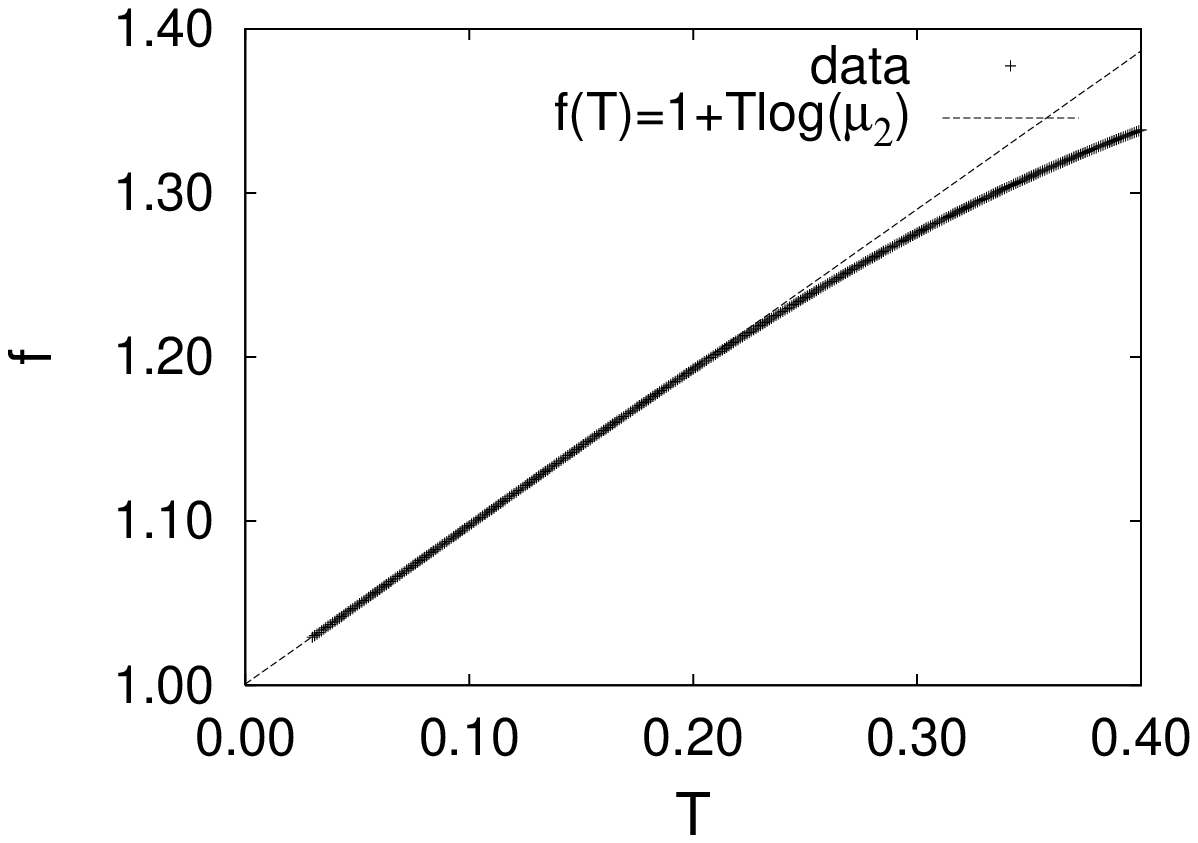}
\caption{\label{fcmu} The critical force for $T\to0$ is found to
fulfill the relation $f(T)=1+T\cdot\log{\mu_{d-1}}$ for both two
dimensions (top) and three dimensions (bottom). The respective surface
connective constants are $\log\mu_2\approx 0.965$ and
$\log\mu_1=0$. The solid line is the data from our simulations and the
dashed line is the relation.}
\end{figure}

There is a simple approximate argument to understand the difference
between both models (the existence of re-entrance in three
dimensions), for details see e.g.\ \cite{orlandini2004,mishra2004}.
Using this argument one finds that for $T$ close to zero in the
$d$-dimensional system the critical force is given by

\begin{equation}
\label{eq_fc}
f^{(d)}_c\approx-\epsilon+T\cdot\log\mu_{d-1}=1+T\cdot\log\mu_{d-1},
\end{equation}
where $\mu_{d-1}$ is the connective constant in $d-1$ dimensions.  The
interesting region of the phase diagram is shown in Fig.\
\ref{fcmu}. Since for two dimensions the entropy for small $T$ is
equal to zero (there are only two configurations contributing at
$T=0$) we see that the critical force is equal to one. In three
dimensions the entropy for small $T$ is equal to the conformational
entropy of self-avoiding walk in two dimensions of length $m_s$.  By
fitting the relation of eqn.\ (\ref{eq_fc}) we find that
$\log\mu_2\approx 0.965$. Given our small system size, this is in
reasonable agreement with the established value of the connective
constant for SAW in two dimensions $\log\mu_2\approx 0.97008\ldots$
\cite{jensen2004}.

\begin{figure}[ht!]
\includegraphics[scale=0.9,angle=0]{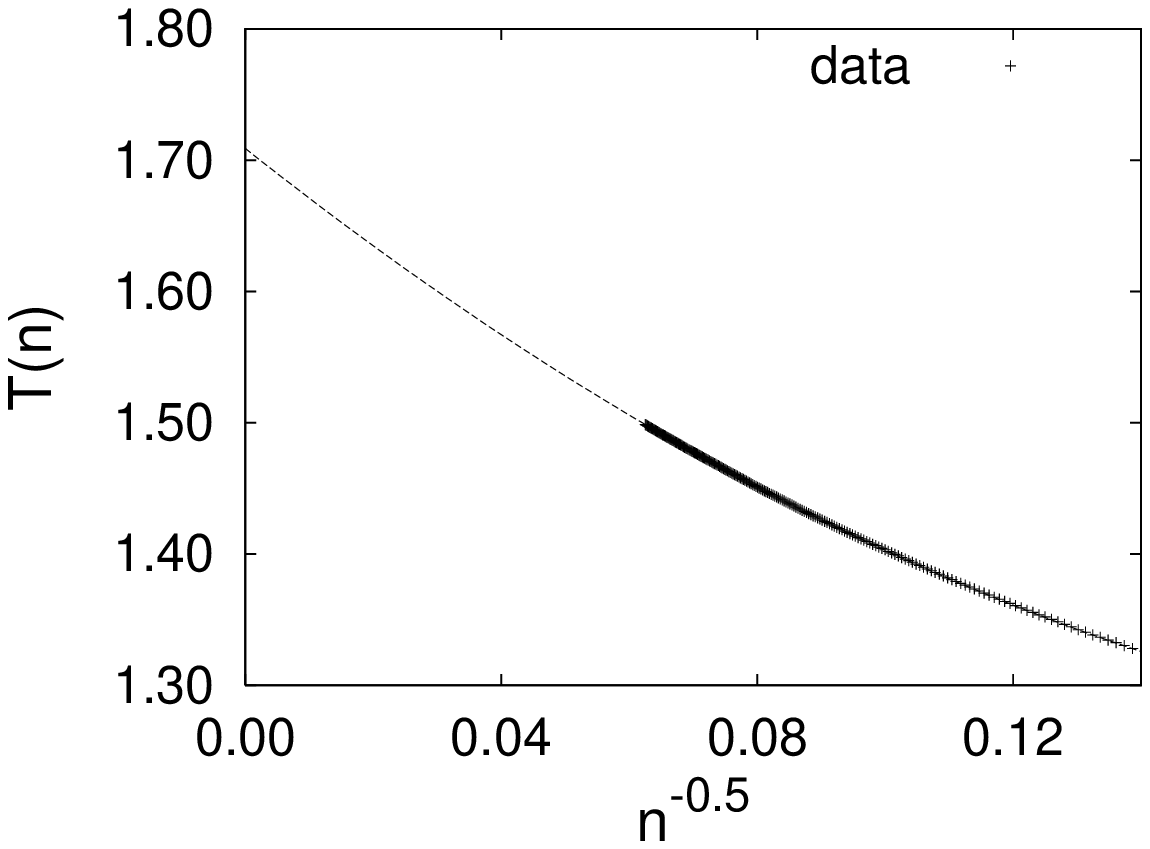}\hfill 
\includegraphics[scale=0.9,angle=0]{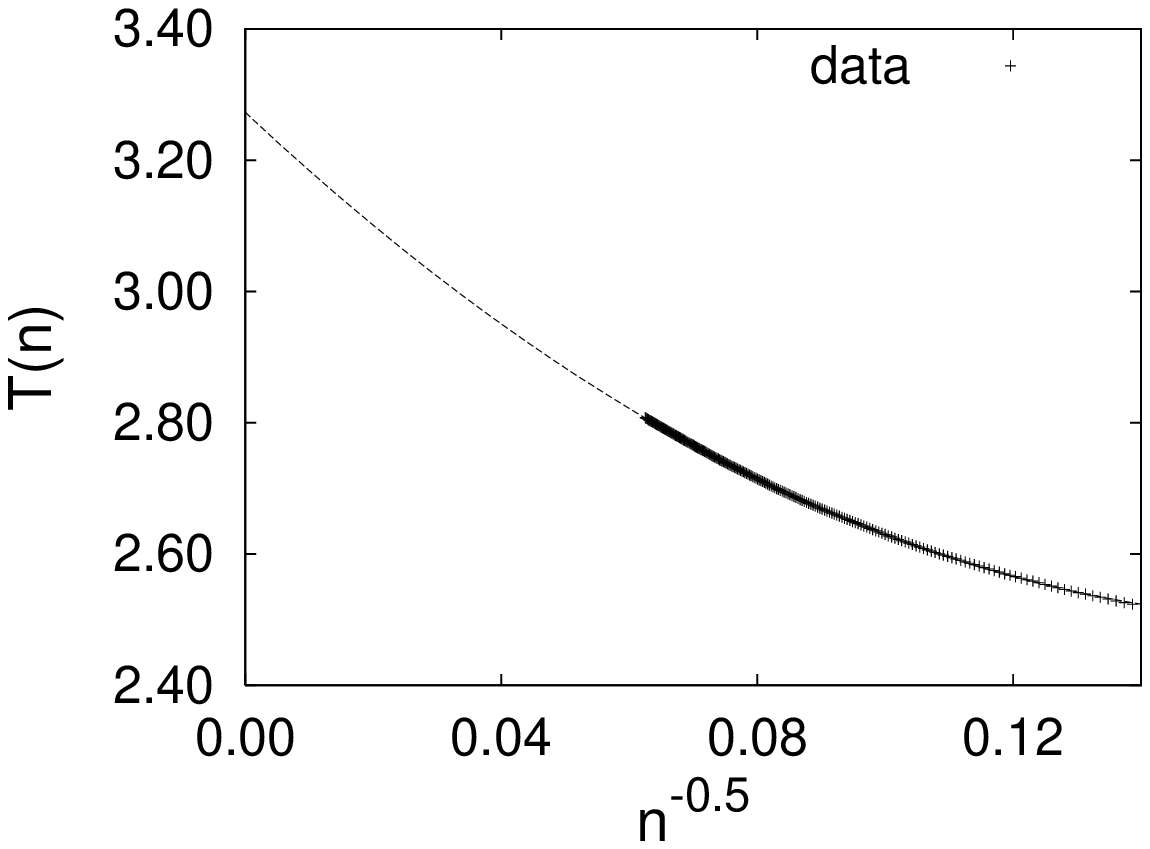}
\caption{\label{fc} The transition temperature between adsorbed and
desorbed phase in the absence of force ($f=0$) for two (top) and three
(bottom) dimensions. The solid line is our data. We approximate the
transition temperature for an infinite system in both dimensions using
a least-squares quadratic fit, shown as a dashed line.}
\end{figure}

If we do not apply any force we have a transition which is driven only
by temperature. The position of this transition depends on $n$ and the
estimates are only approximations of the phase transition location in
the thermodynamic limit. With increasing $n$, the position of the
transition approaches, of course, the real thermodynamic location. In
Fig.\ \ref{fc} we estimate the transition temperature for infinite
systems using the value of the cross-over exponent obtained by
Grassberger and Hegger \cite{grassberger1995} $\phi=1/2$. We see that
the corrections to finite-size scaling are stronger than linear for
both dimensions.  Fitting a simple quadratic function to our data we
find the values $T=1.71$ for two dimensions and $T=3.27$ for three
dimensions. The transition temperature for three dimensions is smaller
than the value found by Vrbova and Prochazka in \cite{vrbova1999}
($T=3.39(2)$) based on simulations of systems of size up to $n=1600$.

\section{Conclusion}
We have presented an application of the flatPERM algorithm to a simple
desorption problem with an intriguing phase-diagram that mimics that
expected for DNA unzipping models. Using flatPERM one may quickly get
a good qualitative overview of the whole phase diagram.  In further
studies we extend our simulations to investigations of adsorption of
interacting self-avoiding walks at a surface, where we have found
intriguing and novel phenomena
\cite{owczarek2004,prellberg2004a}.

\section*{Acknowledgements} 

Financial support from the DFG is gratefully acknowledged by JK and
TP. Financial support from the Australian Research Council is
gratefully acknowledged by ALO and AR. ALO also thanks the Institut
f\"ur Theoretische Physik at the Technische Universit\"at Clausthal.

%


\end{document}